\documentclass{elsart}

\usepackage{amsmath,amssymb}
\usepackage{graphicx}

\begin{document}

\begin{frontmatter}

\title{The SNAP Strong Lens Survey}

\author{Phil Marshall, Roger Blandford and Masao Sako}
\address{Kavli Institute for Particle Astrophysics and Cosmology,\\
P.O. Box 20450, MS29, Stanford, CA 94309, U.S.A.}

\begin{abstract}  
Basic considerations of lens detection and identification indicate that
a wide field survey of the types planned for weak lensing and Type Ia
SNe with SNAP are close to optimal for the optical detection of strong
lenses. Such a ``piggy-back'' survey might be expected even
pessimistically to provide a catalogue of a few thousand new strong
lenses, with the numbers dominated by systems of faint blue galaxies
lensed by foreground ellipticals.  After sketching out our strategy for
detecting and measuring these galaxy lenses using the SNAP images, we
discuss some of the scientific applications of such a large sample of
gravitational lenses: in particular we comment on the partition of
information between lens structure, the source population properties
and cosmology. Understanding this partitioning is key to assessing
strong lens cosmography's value as a cosmological probe.
\end{abstract}

\end{frontmatter}


\section{Introduction}

The proposed SNAP satellite, whilst designed for measuring the
luminosity distance of Type Ia Supernovae as a function of their  
redshift~\cite{COS/Per++99,COS/Rie++04} and probing the
large-scale structure in the Universe through weak gravitational
lensing~\cite{COS/Rho++01}, will be an observatory offering
high angular resolution optical imaging (PSF FWHM $\sim 0.12$ arcsec)
over a wide survey field (1000 square degrees) in many (9)
filters~\cite{SNAP}.
This potent combination satisfies the basic requirements of an optical
strong gravitational lensing survey. Experience with radio surveys such
as CLASS~\cite{GL/Mye++03} suggests that around one in 1000 high redshift objects
may be strongly lensed, whilst the lensing rate for the less visible
but far more numerous faint optical galaxies in the optical appears to
be significantly lower. These numbers prompt for the
examination of very large areas of sky. However, the majority of the
multiple-imaging events occur on arcsecond angular scales, requiring
sub-arcsecond resolution for their discovery.  Finally, the
identification of a strong lens system hinges on the achromatic nature
of lensing: multi-colour imaging is of enormous value in the initial
identification of a lens.

The lensing cross-section in the Universe is dominated by the many
massive elliptical galaxies lying at redshifts of 1 or 
less~\cite{GL/TOG84}; galaxy clusters provide more cross-section per object but
are rarer.  The following strategy suggests itself: search in the
vicinity of (easily classified) elliptical galaxies, with
(well-understood) photometric redshifts, for characteristic
patterns of similarly-coloured images. Similar algorithms have been
applied to spectroscopic data~\cite{GL/Bol++04} and, principally
visually, to HST image data~\cite{GL/Rat++99,GL/Fas++04}.  Many lens
systems may be expected from simply filtering the 9 filter  photometric
catalogue from the weak lensing analysis; others will be found by
digging deeper into the wings of the lens galaxy light. We begin to
quantify the success of SNAP in this regard below.
Cluster gravitational lenses are perhaps best discovered by a similarly
targeted search for elongated images around congregations of red
sequence   galaxies~\cite{GL/LSS04,GL/Gla++03}, followed by an
iterative search for more multiple images. Whilst forming part of the
ongoing SNAP strong lensing survey project, these lens systems are not
discussed here. Instead, we focus on the elliptical galaxy lenses and
some of the science they enable.


\section{Predicting elliptical lens numbers}

\begin{table}[t]
\centering
\begin{tabular}{lcccc}
\hline
Survey	& $\Omega$ / sq.\ degrees	& AB magnitude limit	& $N_{\rm lens, galaxies}$	& $N_{\rm lens, quasars}$	\\
\hline\hline
Deep (SNIa)	& 15					& 30.6			& 5000	& few--few tens	\\
Wide (WL)	& 1000				& 28.3			& 50000	& 100--1000	\\
\hline
\end{tabular}
\medskip
\caption{Proposed primary SNAP surveys, and projected elliptical strong
gravitational lens numbers. The magnitude limits are for point sources in the B band (filter 1).}
\label{tab:surveys} 
\end{table}

The leftmost columns of Table~\ref{tab:surveys} give the vital
statistics of the two baseline surveys planned for SNAP's initial
4-year observing period. In this work we investigate the capabilities
of a ``piggy-back'' survey,  just using the images taken for the two
primary experiments and placing no extra demands on the mission.

The expected numbers of lenses are given by the following equations:
\begin{align}
&N_{\rm lens} = \int X \cdot \frac{d^2 N_{\rm d}}{dz_{\rm d} d\sigma_{\rm d}} 
\cdot \frac{d^2 N_{\rm s}}{dz_{\rm s} dm_{\rm s}}\,\, dz_{\rm d}\, 
d\sigma_{\rm d}\, dz_{\rm s}\, dm_{\rm s}
\label{eq:numbers}
\\
&X(z_d,\sigma_{\rm d},z_{\rm s},m_{\rm s}) = \int^{\boldsymbol{\beta}_{\rm crit}} 2 \pi S(\boldsymbol{\beta},...)\,\, d^2\boldsymbol{\beta}
\label{eq:xsection}
\end{align}
We follow Mitchell et al.~\cite{GL/Mit++04} and use the SDSS elliptical
galaxy velocity function for the deflector population 
$d^2 N_{\rm d} / dz_{\rm d} d\sigma_{\rm d}$; 
for the faint galaxy sources 
($d^2 N_{\rm s} / dz_{\rm s} dm_{\rm s}$)
we extrapolate somewhat conservatively the
HDF galaxy counts from Casertano et al.~\cite{O/Cas++02} and apply the model redshift distribution of Massey et al.~\cite{GL/Mas++04}. For
the quasar
sources we use the 2dF
luminosity functions from Boyle et al. (2000) and Croom et al. 
(2004)~\cite{O/Boy++00,O/Cro++04}, 
noting that these analyses give different faint end slopes. 
The final piece of equation~\ref{eq:numbers} is the
cross-section for producing an observable multiple image system. This
quantity is an integral over the source plane, and is dependent on the
lens properties (assumed for simplicity to be singular isothermal
spheres~\cite{GL/Koo++03}), but also the details of the detection
process.  A multiple image system was taken as having been detected if
three pixels of the surface brightness peaks were three times higher
than the noise level in the filter~1 image, provided the images were
separated by twice the PSF FWHM.
\begin{figure}[!t]
\begin{minipage}[t]{0.48\linewidth}
\centering\includegraphics[width=\linewidth]{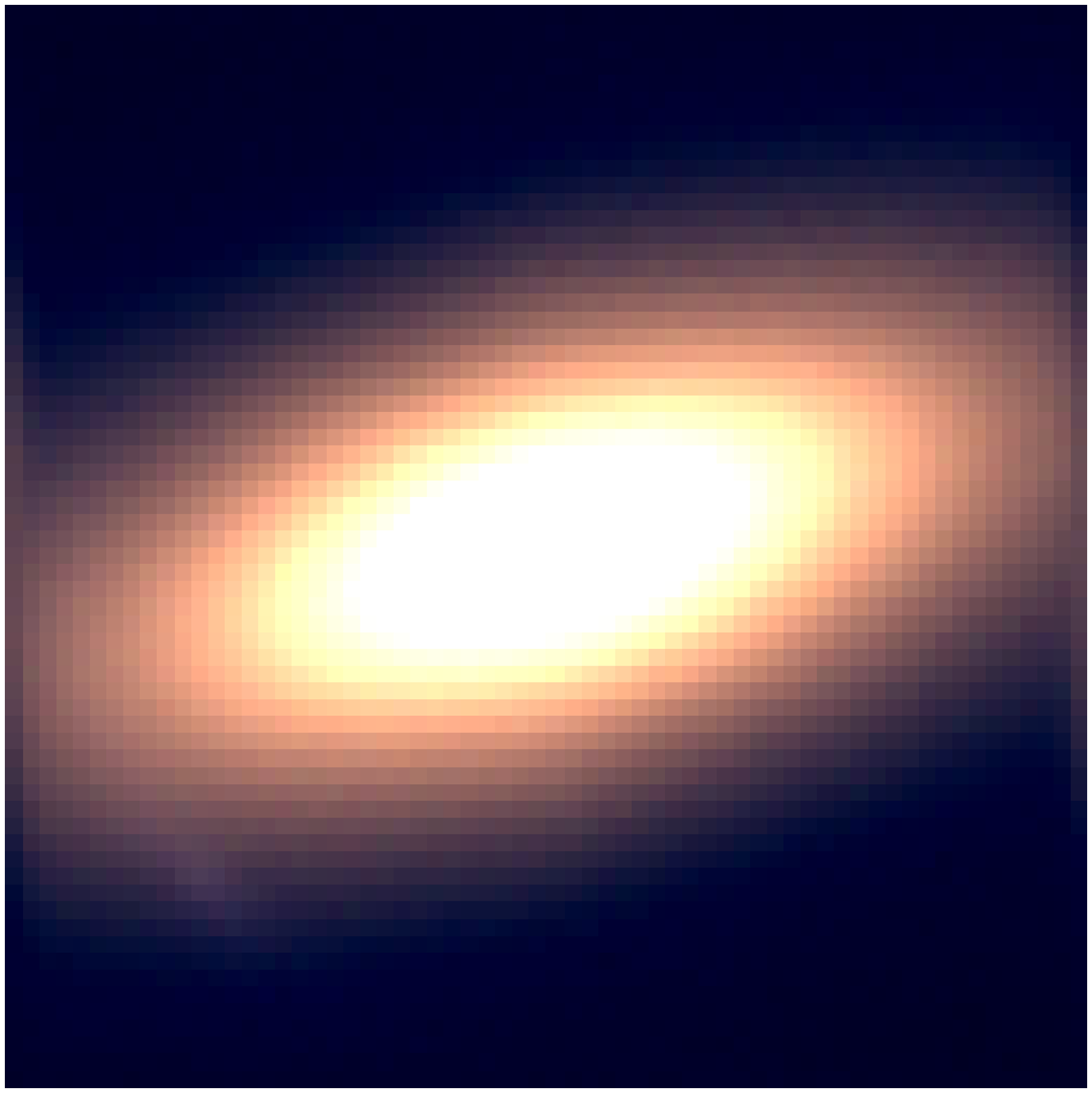}
\end{minipage} \hfill
\begin{minipage}[t]{0.48\linewidth}
\centering\includegraphics[width=\linewidth]{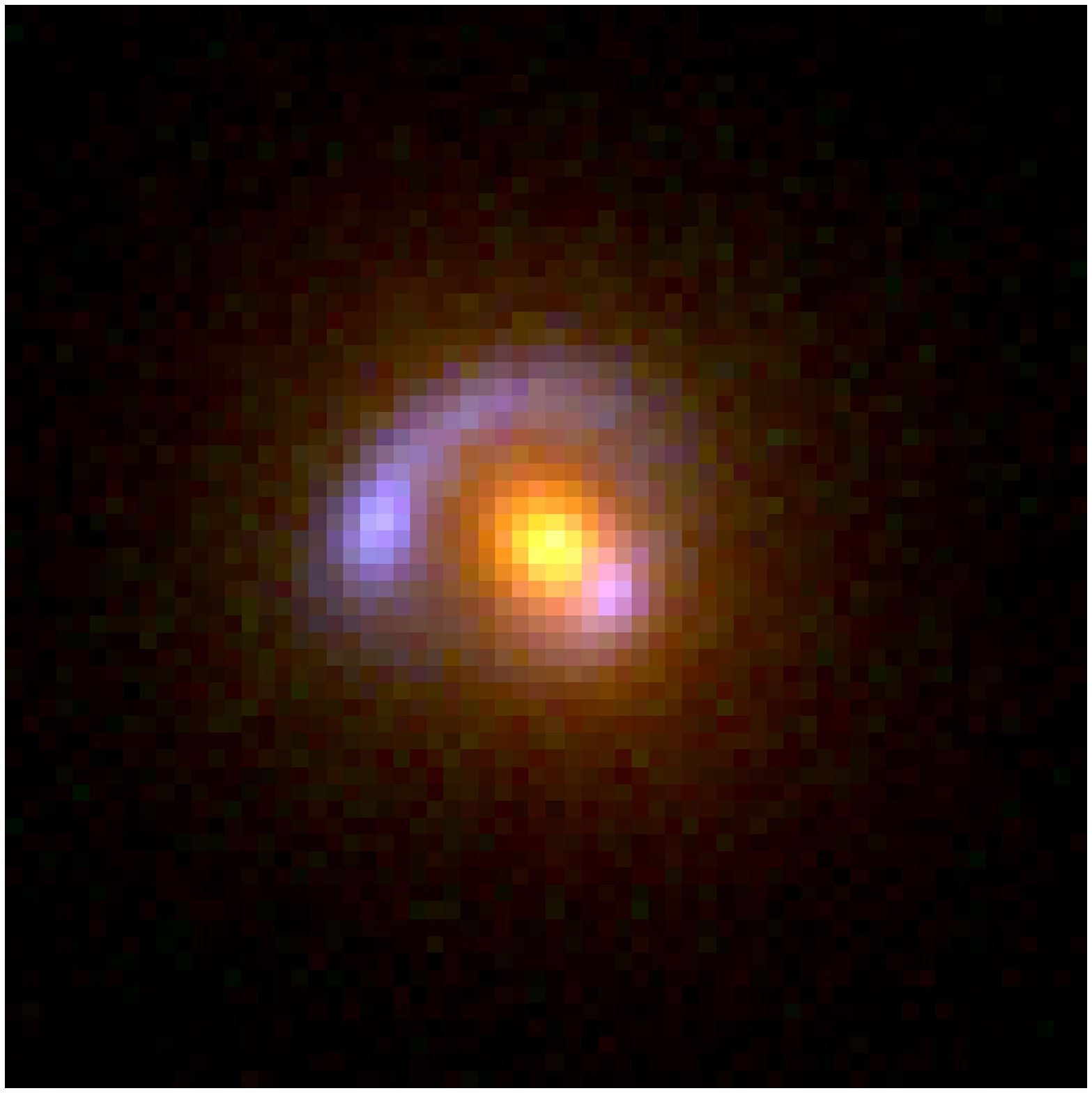}
\end{minipage}
\caption{Simulated typical strong lens systems, as they would be 
observed with SNAP
(these images are 9 filter composites). The fields are centred on the
(yellow) elliptical lens galaxy; the lensed background source galaxies
appear as (blue) arcs with $\sim 1$ arcsecond curvature radii.}
\label{fig:images}
\end{figure}

The results of this simple calculation are shown in the righthand
columns of Table~\ref{tab:surveys}: they suggest that around 1-2\% of
elliptical galaxies have a multiple image system associated with
them. The lensed quasar numbers are most sensitive to the (as yet somewhat poorly
measured) faint-end slope of the luminosity function. The galaxy-galaxy
lens count prediction is quite sensitive to our ability to separate the
lens galaxy light from the source image: the above selection function
made no allowance for the lens galaxy. An approximate actual 
success rate was
obtained by visual inspection of a small sample of lenses drawn from
the probability distribution defined by the integrand in
equation~\ref{eq:numbers}. Figure~\ref{fig:images} gives an example of
a lens system measurable with SNAP, and a system that may well have
escaped detection by eye; in these images, the 9 filters have been
crudely grouped in threes for the colourisation, and a square root
stretch applied. Our preliminary estimate is that as many as 1 in 5
galaxy-galaxy strong lens systems expected from the ``invisible lens''
selection function will actually be detectable and measurable in the
SNAP images; 
the quasar lenses will be much easier to detect, but their numbers ae still uncertain by an order of magnitude or so.
We may hope to improve the galaxy-galaxy lens detection
 rate with sophisticated
9-filter lens finding algorithms, but for now we anticipate a catalogue of some 10000 strong lenses, dominated by galaxy-galaxy lenses detected in the weak lensing survey.
  

\section{Science from the elliptical lens sample}

\begin{figure}[!t]
\begin{minipage}[t]{0.43\linewidth}
\centering\includegraphics[width=\linewidth]{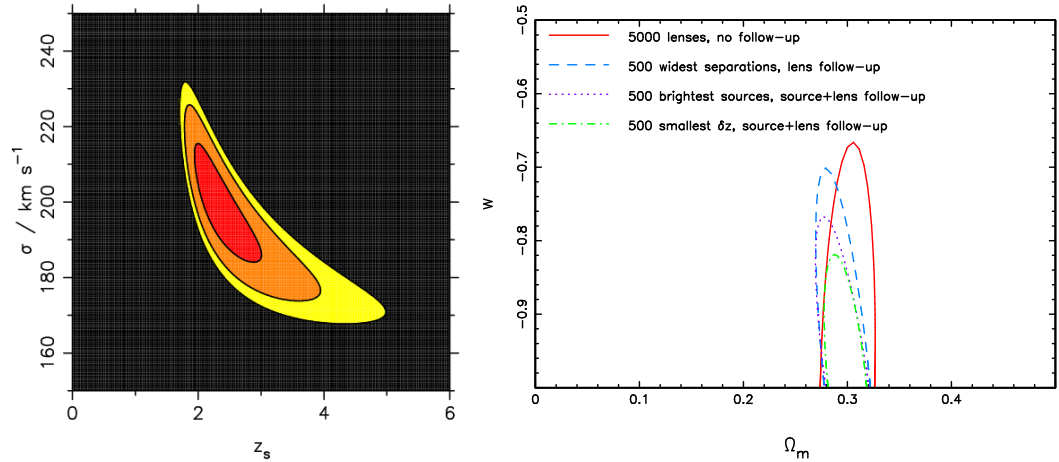}
\end{minipage} \hfill
\begin{minipage}[t]{0.55\linewidth}
\centering\includegraphics[width=\linewidth]{figure2b.ps}
\end{minipage}
\caption{Left: source redshift measurement 
($z_{\rm s}=2.3$) from a $z_{\rm d}=1.1$ lens. 
Right: strong lens cosmography from various, and relatively unambitious, subsamples of a 
SNAP catalogue.}
\label{fig:partition}
\end{figure}

The expected SNAP lens sample
would contain much information on the elliptical galaxy mass
distributions, the source redshifts and cosmology. How is the information
partitioned? An approximate answer to this can be found by
qualitatively identifying the subsets of the lens catalogue with 
discriminating power in each field.  In the lefthand panel of
Figure~\ref{fig:partition} we show how the redshift distribution of 
galaxies too faint for spectroscopy can begin to be investigated using a snapshot
spectroscopic follow up campaign on the candidate lens galaxies
(providing knowledge of the lens mass and lens redshift). Those systems
where the lens and source planes are closely separated give the
tightest constraint on the source redshift.   
Conversely, a subsample of the SNAP strong lens catalogue possessing
well-understood source redshifts would open up the possibility of a
cosmographic study~\cite{GL/P+G81,GL/IGR97}.  The image separation is dependent on cosmology,
being proportional to the ratio of angular diameter distances 
$D_{\rm ds}(z_{\rm d},z_{\rm s})/D_{\rm s}(z_{\rm s})$.

The righthand panel of Figure~\ref{fig:partition} shows the constraints
available in the $\Omega_{\rm m} - w$ plane (assuming a flat Universe)
from various cosmographic experiments. A conservative estimate of 5000
suitable lenses (with image separations detected and measured to 1/3 of
a pixel precision), with photometric redshift uncertainties of 0.02
(lens) and 0.2 (source) and the lens mass just estimated from the SDSS
Faber Jackson relation~\cite{O/Ber++03} yields interesting, but not
competitive, precision. Tighter constraints are gained by focussing on
sub-samples of the catalogue 
and targeting them for spectroscopic follow-up observation. Constraining
the mass with spectral velocity dispersion measurements has the biggest
effect, suggesting that the SNAP data alone are best used to
investigate the mass distribution of elliptical galaxies within an
assumed cosmological model. The next step is the
assessment of the effect of the lens modelling on the dark energy parameter accuracy. 

As seen above, SNAP will detect a significant number of lensed quasars
and provide HST-quality optical imaging data for these objects,
promising to greatly improve the strong lensing measurment of $H_0$.
What will be needed are complementary time delay measurements: the deep
SN survey (with its 4-day observing cadence) could provide time delays
for some tens of lensed quasar systems, but the much larger weak
lensing survey will only provide single epoch images. There will be a
number of survey telescopes operational at the same time as SNAP that
could provide time delays for these latter objects; these, together
with the SNAP imaging data, should provide excellent constraints on the
Hubble constant and allow its scatter (which is sensitive to
small-scale mass fluctuations along the line-of-sight) to be
investigated.  

\section{Conclusions}

The width and depth of the proposed SNAP observing programmes, their
multi-colour nature, and the excellent image quality are all
well-suited to a strong lens search. Targeting the elliptical galaxies,
we believe that the achromatic excesses in the images can be extracted;
the resulting catalogue, containing an unprecedented number of multiple
image systems, may then be used for many astrophysical and cosmological
experiments, with just a few touched upon here. We anticipate that
complementary observations in the spectral and temporal domains will be
very important for getting the most out of the SNAP data. 


\section{Acknowledgments}

We thank Dragan Huterer and Charlie Baltay for useful discussions.
This work was supported by the Department of Energy contract DE-AC3-76SF00515 to SLAC.


\end{document}